\documentclass[12pt,draftclsnofoot,onecolumn]{IEEEtran}
\pdfoutput=1

\usepackage{cite}
\usepackage{graphicx}
\usepackage[justification=centering]{caption}
\usepackage{autobreak}
\usepackage{setspace}
\usepackage{amsmath}
\usepackage{booktabs}

\ifCLASSINFOpdf
\else
\fi
\begin{document}
%
\title{Retrospective Interference Regeneration Schemes for Relay-Aided $K$-user MIMO Downlink Networks}

\author{
        Jingfu~Li,~\IEEEmembership{}
        Zehui~Xiong,~\IEEEmembership{Member,~IEEE},
       Dusit~Niyato,~\IEEEmembership{Fellow,~IEEE},
       Weifeng~Su,~\IEEEmembership{Fellow,~IEEE},
      Wenjiang~Feng,~\IEEEmembership{} and
        Weiheng~Jiang,~\IEEEmembership{Member,~IEEE}  
}

\maketitle

\vspace{-1.5cm}

\setlength{\footnotesep}{0.2cm}
\captionsetup{aboveskip=10pt}
\captionsetup{belowskip=-10pt}

\begin{abstract}
To accommodate the explosive growth of the Internet-of-Things (IoT), incorporating interference alignment (IA) into existing multiple access (MA) schemes is under investigation. However, when it is applied in MIMO networks to improve the system compacity, the incoming problem regarding information delay arises which does not meet the requirement of low-latency. Therefore, in this paper, we first propose a new metric, degree of delay (DoD), to quantify the issue of information delay, and characterize DoD for three typical transmission schemes, i.e., TDMA, beamforming based TDMA (BD-TDMA), and retrospective interference alignment (RIA). By analyzing DoD in these schemes, its value mainly depends on three factors, i.e., delay sensitive factor, size of data set, and queueing delay slot. The first two reflect the relationship between quality of service (QoS) and information delay sensitivity, and normalize time cost for each symbol, respectively. These two factors are independent of the transmission schemes, and thus we aim to reduce the queueing delay slot to improve DoD. Herein, three novel joint IA schemes are proposed for MIMO downlink networks with different number of users. That is, hybrid antenna array based partial interference elimination and retrospective interference regeneration scheme (HAA-PIE-RIR), HAA based improved PIE and RIR scheme (HAA-IPIE-RIR), and HAA based cyclic interference elimination and RIR scheme (HAA-CIE-RIR). Based on the first scheme, the second scheme extends the application scenarios from $2$-user to $K$-user while causing heavy computational burden. The third scheme relieves such computational burden, though it has certain degree of freedom (DoF) loss due to insufficient utilization of space resources. Overall, our proposed schemes are able to solve the issue of information delay caused by IA techniques and further improve system capacity on the basis of RIA scheme.
\end{abstract}

\begin{IEEEkeywords}
K-user MIMO downlikn networks, degree of freedom, degree of delay, hybrid antenna array structure of relay, interference elimination and retrospective interference regeneration.
\end{IEEEkeywords}

%
\IEEEpeerreviewmaketitle

\clearpage

\section{Introduction}
%
%
%
%
\IEEEPARstart{A}{s} widely known that, for modern cellular mobile communication networks, the multiple access (MA) technology, as one of the key technologies, always plays an import role in improving the system capacity and supporting multiple users to share the same cellular resources over the same region. In the past decades, various MA technologies have been proposed, e.g., TDMA, FDMA, CDMA, and OFDMA. Recently, with the explosive growth of wireless connection demands, the innovation of MA technology is extremely urgent and two aspects for improving the efficiency of MA are being considered. One is to improve dimensions of orthogonal space, such as the spatial division multiple access (SDMA) proposed in \cite{681322}, where spatial beamforming is used to achieve spatial gain. However, its performance is limited by the challenge of omnidirectional antenna structure. While if the smart beamforming directional antennas \cite{2021Joint,2021Joint2} are used, i.e., for the massive MIMO system, its spatial gain can be further improved. Another way of improving MA is to increase the resolution among signals. To achieve this goal, various of non-orthogonal multiple access (NOMA) techniques are introduced, e.g., power division-based NOMA (PD-NOMA) \cite{8352617}, sparse code multiple access (SCMA) \cite{7510759}, pattern division multiple access (PDMA) \cite{7752716}, and multiple users sharing access (MUSA) \cite{8891156}. Among them, PD-NOMA and SCMA are the promising schemes for the fifth-generation system (5G) \cite{7414384}. The former adopts superposition coding with the transmit powers to shares the radio spectrum as weight factors, and the latter maps symbols to sparse codes.

Now, with rapid development of the 5G \cite{9174806} and the evolution from the Internet-of-Things (IoTs) to the Internet-of-Everything (IoE) \cite{9205230}, the services and applications, such as industry 4.0 and robotics \cite{9122412}, whose requirements of high capacity to connect more IoE terminals as well as low latency for their services, are presented and dramatically increased. Unfortunately, though many efforts have been devoted in this filed, we still lack effective technologies to support the significantly increasing system connection demands, not to mention their low latency requirements. At the same time, another technique named interference alignment (IA) \cite{6594784} has been attracting attention during the past few years. It utilizes channel state information at the transmitter (CSIT) to reconstruct the structure of signals, where the interference casts overlapping shadows at the unintended receivers. From the study of IA, we know that its performance, e.g., degree of freedom (DoF) \cite{6831204}, is restricted by the demand of instantaneous CSIT. It means that, when the CSIT feedback from receivers to transmitters experiences an unavoidable delay, the acquired CSIT cannot be used at transmitter sides. To handle this issue, retrospective interference alignment (RIA) with delayed CSIT is studied \cite{Maddah2012Completely, 2011Achieving, 2014On, Hao2016Achievable,6111234, 2015Retrospective, 2015Retrospective2,6530617, 6874856, 7605466}. \cite{Maddah2012Completely} first introduces the RIA scheme for the MISO scenario. It leverages the repetition coding $R$ times in the first phase and then, with interference information of previous slots, each receiver achieves jointly decoding in the second phase. Subsequently, the discussion about RIA is extended to other scenarios with different system configurations, i.e., interference channel (IC) \cite{2011Achieving, 2014On, Hao2016Achievable}, broadcast channel (BC) \cite{6111234, 2015Retrospective, 2015Retrospective2}, and X-channel (XC) \cite{6530617, 6874856, 7605466}. In these studies, the DoF can achieve a considerable gain but has not reached its upper bound.

From existing studies, we can see that IA techniques have been discussed in different scenarios and related works are also increasing \cite{8600208, 9145529, 9217497,2021Two}. However, these studies focus only on the DoF performance, i.e., to enhance the capacity of the system for serving as many users as possible with the same resource, while no studies concern the latency issue, which is the most important performance measure for the IoE service and next generation networks. Taking the RIA as an example, during the transmission, no signal can be decoded at the receiver until the second phase. Its delayed decoding will cause serious problems for time-sensitive industrial networks \cite{9187995}. To cope with this issue and to improve DoF for the IA-assisted networks, in this paper, we first propose the degree of delay (DoD) metric. It specifies the average information delay for each symbol with a comprehensive consideration of time span and priority. Then, for MIMO networks with BC, we introduce three kinds of novel IA schemes, i.e., hybrid antenna array based partial interference elimination and retrospective interference regeneration scheme (HAA-PIE-RIR), HAA based improved PIE and RIR scheme (HAA-IPIE-RIR), and HAA based cyclic interference elimination and RIR scheme (HAA-CIE-RIR). The first one is mainly used in the $2$-user MIMO scenarios and the latter two are designed for $K$-user MIMO scenarios. Under the conditions of different number of users, each of them can reduce DoD and meanwhile improve DoF. The main contribution is five-fold as follows:

\begin{itemize}
\item For BC networks, DoD is proposed to characterize the degree of information delay caused by transmission schemes, and to incorporate the priority of traffic, which is based on time-delay sensitivity, into account. Then, under three kinds of typical transmission schemes, i.e., time division multiple address scheme (TDMA), RIA scheme, and beamforming based distributed TDMA scheme (BD-TDMA), the influence factors of DoD are analyzed. We analytically prove that DoD depends on delay sensitive factor $DSF$, size of data set ${Num}$, and queueing delay slot ${n_s}$.
\item To reduce DoD and obtain a higher DoF, the HAA-PIE-RIR scheme is presented for $2$-user $M \times N$ BC networks. Accordingly, the antenna array structure with omnidirectional and directional antennas is designed at the relay side to eliminate IUI and erase redundant symbols of desired signals. Meanwhile, distributed retrospective interference regeneration (DRIR) is adopted to compensate missing symbols which are erased before. The performance of the proposed scheme is superior to that of RIA scheme in DoD and DoF.
\item Since HAA-PIE-RIR scheme is hard to be extended to multi-user scenarios, HAA-IPIE-RIR scheme is then presented for $K$-user $M \times N$ BC networks. It redesigns the precoding matrix of relay where IUIs of multiple users can be eliminated simultaneously. Thus, the performance of the proposed scheme is further improved while it causes considerable computational burden to the relay,  which makes it not applicable for the case that the number of transceiver antennas is large.
\item To reduce computational complexity of precoding matrix design, we propose the HAA-CIE-RIR scheme. The scheme takes a circular transmission strategy to simplify the scenario of $K$-user $M \times N$ BC networks into the scenario of $2$-user $M \times N$ BC networks. Meanwhile, DRIR algorithm is applied to compensate missing symbols. Compared with the HAA-IPIE-RIR scheme, this scheme relieves the pressure from precoding matrix design while losing some DoF gain.
\item For $K$-user $M \times N$ BC networks, performances of the proposed schemes are extensively evaluated in three aspects, namely DoD, DoF and computational cost. 	The simulation results demonstrate that HAA-IPIE-RIR scheme obtains the optimal DoD and DoF while the computational cost is relatively high. By contrast, HAA-CIE-RIR scheme decreases computational cost but experiences certain performance loss in DoD and DoF. In general, the performances of three proposed schemes are superior to that of RIA scheme, especially in terms of DoD and DoF.
\end{itemize}

The rest of this paper is organized as follows. In Section II, we present the $K$-user MIMO BC networks model, the CSIT feedback model and also the definition of the DoF. In Section III, the definition of DoD is introduced and we analyze the DoD performance for three typical transmission schemes. To improve the DoF and reduce the DoD for the considered networks, three novel IA schemes are proposed and discussed in Section IV and Section V, respectively. The performance of the proposed schemes are analyzed by simulations in Section VI and we conclude at last in Section VII.

\section{System Model}
We consider the $K$-user MIMO BC network \cite{5399360} as shown in Fig. 1. The configuration of the network is set to $(K,M,N)$, in which one base station with $M$ antennas serves $K$ users and each user is equipped with $N$ antennas. Define the sets of users, transmitting antennas, receiving antennas, and time slots as ${\boldsymbol{K}}{ = }\{ 1,2, \ldots ,K\}$, ${\boldsymbol{M}}{ = }\{ 1,2, \ldots ,M\}$, ${\boldsymbol{N}}{ = }\{ 1,2, \ldots ,N\}$, and ${\boldsymbol{T}}{ = }\{ 1,2, \ldots ,T\}$, respectively. Due to the broadcast nature of the wireless communication, when the base station transmits signals to user $k,\forall k \in {\boldsymbol K}$, other users $- k = {\boldsymbol K}\backslash k$ also receive signals from the base station. We assume that all users share the same spectrum resources and the base station serves multiple users simultaneously, and thus the base station will cause IUI to all users. For ease of presentation, we divide transmitting antennas into clusters and define clusters set as ${\boldsymbol{I}}{ = }\{ 1,2, \ldots ,K\}$. For the first $K-1$ clusters, each cluster is allocated with $\left\lceil {M/K} \right\rceil$ antennas where $\left\lceil {\rm{*}} \right\rceil $ denotes round up operation, and the last cluster is allocated with $M-\left\lceil {M/K} \right\rceil$ antennas. Then, the component of received signals from cluster $i,\forall i \in {\boldsymbol I}$ to user $k,\forall k \in {\boldsymbol K}$ has two parts. The first part is desired signals where $i=k$, and the second part represents IUI. Herein, details of communication process can be described in time domain \cite{5358700}.
\begin{figure}[h]
\centering
\includegraphics[width=4in]{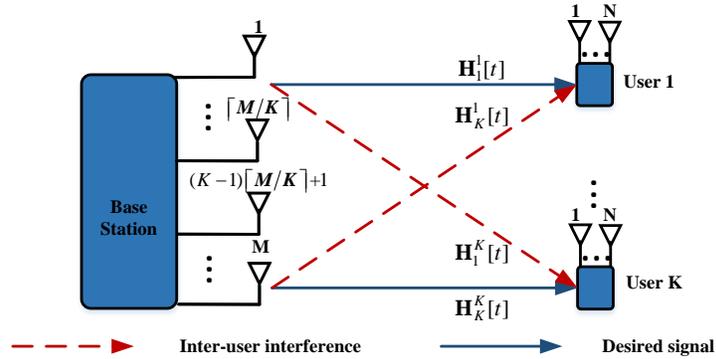}
\caption{$K$-user MIMO broadcast channel network.}
\label{Fig1}
\end{figure}
\vspace{-1.0em}

\subsection{Time Domain Model}
Without loss of generality, in the following, we take one transmission slot $t,\forall t \in {\boldsymbol{T}} $ in $K$-user MIMO BC network as an example. In this scenario, each cluster $i,\forall i \in {\boldsymbol I}$ of the base station transmits symbols ${\boldsymbol{{S}}_{i}}[t]$ with BC. The received signal of user $k$ is written as
\begin{equation}
\label{eq1}
{\boldsymbol{{y}}_{k}}[t]{ = }{\sum\limits_{i = 1}^K {{\bf{H}}_i^{k}[t]{{\boldsymbol{{S}}}_{i}}[t]} } { + }\overline {\boldsymbol{{N}}},
\end{equation}
where ${\bf{H}}_i^{k}[t] \in{\boldsymbol{C}}{^{N \times \left\lceil {M/K} \right\rceil }}$ denotes the channel matrix between cluster $i$ of the base station and user $k$, which is independent and identically distributed $\left( {i.i.d.} \right)$ according to ${\cal C}{\cal N}(0,1)$. $\overline {\boldsymbol{N}}  \in{\boldsymbol{C}}{^{N \times 1}}$ denotes the additive white Gaussian noise (AWGN) term at user $k$.

Based on the model, we know that each user $k$ is equipped with $N$ antennas to receive signals from base station's $M$ antennas. Then, we assume that $M>N$, which stands for the fact that desired symbols ${\boldsymbol{{S}}_{i}}[t]$ cannot be decoded within only one slot and additional transmissions are required. To handle this issue, IA techniques are introduced and corresponding CSI feedback model is indispensable.

\subsection{CSIT Feedback Model}
To make IA applicable in the considered scenario, CSI acquisition is requisite. However, when transmitter receives CSI from receiver by feedback link, the CSI experiences an unavoidable feedback delay \cite{7822959}. Thus, the obtained CSI is a delayed version, which means that the channel matrix may have changed. It brings a potential problem that, if transmitter designs precoding matrix with previous channel matrix information to eliminate interference of current slot, it cannot be achieved. For this problem, the CSIT feedback model \cite{6926832} is introduced, from which different types of CSI are separated and each type is used in different way. The specific classifications of CSI are shown in Fig. 2.
\begin{figure}[h]
\vspace{-2.0em}
\centering
\includegraphics[width=3.5in]{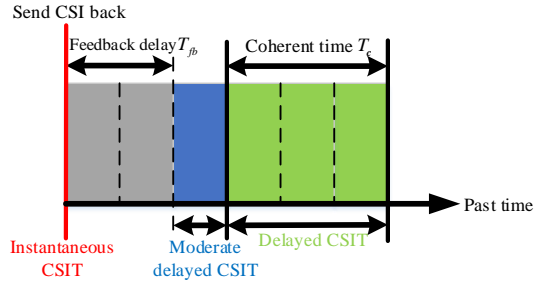}
\caption{CSIT feedback model.}
\label{Fig2}
\end{figure}

We assume that the CSI is an infinite accuracy CSI \cite{6117048}, i.e., without quantization error. Then, we only need to discuss the relationship between CSI and feedback delay. We define the channel feedback delay and the period of coherence time as ${T_{fb}}$ and ${T_c}$, respectively. The definition of ratio $\lambda$ between ${T_{fb}}$ and ${T_c}$ is given by
\begin{equation}
\label{eq2}
\lambda { = }{{{T_{fb}}} \mathord{\left/
 {\vphantom {{{T_{fb}}} {{T_c}}}} \right.
 \kern-\nulldelimiterspace} {{T_c}}}.
\end{equation}
According to the value of $\lambda$, CSI is classified into three types and the meaning of each type as:

\begin{itemize}
\item $\lambda  = 0$, it represents instantaneous CSI, i.e., when the receiver finishes channel estimation, the transmitter obtains the estimated channel state information immediately, namely without feedback delay. Therefore, the transmitter can design precoding matrix with instantaneous delayed CSI to eliminate interference of the current slot. However, this situation is too ideal to be achieved.

\item $0 < \lambda  < 1$, it should be further discussed with two subcases, i.e., the CSI feedback delay is in the interval of $\left[ {{T_{fb}} ,{T_c}- {T_{fb}}} \right]$ or $\left[ {{T_c} - {T_{fb}},{T_c}} \right]$. For the former, it is equivalent to the case of the instantaneous CSI. For the latter, it is the case of moderately delayed CSI, which means that, the feedback delay exists but the channel matrix does not change in the remaining time of the coherent time. Thus, the transmitter can design precoding matrix with moderately delayed CSI to eliminate interference of current slot.

\item $\lambda  \ge 1$, it represents delayed CSI which is an absolutely outdated version. In specific, when the transmitter obtains CSI, the past time exceeds the relevant time so that the channel matrix has already changed. Under this situation, if we take delayed CSI to conduct the precoding matrix design, the interference of the current slot cannot be eliminated completely with IA technique. However, with further analysis, we find that both delayed CSI and the channel matrix of past slots are determinate and fixed. In other words, delayed CSI can be used for precoding matrix design of current slot to eliminate the interference of past slots.
\end{itemize}
In our proposed schemes, we take a combined CSI, i.e., associate moderately delayed CSI with delayed CSI. As such, we can conduct the precoding matrix design to eliminate interference of each slot. The specific application will be presented in Section IV and Section V.

\subsection{Sum Degrees of Freedom}
To quantify the performance of the proposed schemes, the sum degrees of freedom \cite{7442513} is used in this paper. Without loss of generality, for each slot $t,\forall t \!\in\!{\boldsymbol{T}}$, let $\{{{\bf{S}}_{1}}[t], \ldots ,{{\bf{S}}_{k}}[t]\}$ and $\{ {R_{1}}, \ldots ,{R_{k}}\}$ denote the transmitting signals and achievable rates for the cluster $\{ 1,2, \ldots ,K\}$ of the base station, respectively. Thus, for user $k,\forall k \in {\boldsymbol{K}}$, its achievable rate is expressed by
\begin{equation}
\label{eq3}
{R_{k}}{ = }{d_{k}}\log (1{ + }SNR) + o\left( {\log (SNR)} \right),
\end{equation}
where ${d_{k}}$ denotes the DoF of the user $k$ and its definition is expressed by
\begin{equation}
\label{eq4}
{d_{k}} = \mathop {\lim }\limits_{SNR \to \infty } {{{R_{k}}} \mathord{\left/
 {\vphantom {{{R_{k}}} {{{\log }_2}(SNR)}}} \right.
 \kern-\nulldelimiterspace} {{{\log }_2}(SNR)}}.
\end{equation}
For interference networks, the DoF is extended to the sum-DoF which is calculated by
\begin{equation}
\label{eq5}
DoF_{sum} = \mathop {\lim }\limits_{SNR \to \infty } \mathop {\max }\limits_{\boldsymbol{R}}  {\sum\limits_{k = 1}^K {{{{R_{k}}} \mathord{\left/
 {\vphantom {{{R_{k}}} {{{\log }_2}(SNR)}}} \right.
 \kern-\nulldelimiterspace} {{{\log }_2}(SNR)}}} },
\end{equation}
where ${\boldsymbol{R}}{ = }\left[ {{R_{1}}, \ldots ,{R_{k}}} \right] \in {{\boldsymbol{R}}^K}$ denotes the vector of achievable rates for all users. Herein, for our considered network, $DoF_{sum}$ is capable of measuring the system capability and, for ease of presentation, we substitute $DoF_{sum}$ with $DoF$ in the rest of the paper.

\section{Time Delay Evaluation Criterion}
As mentioned earlier, the studies about DoF and IA techniques have attracted considerable attention \cite{Maddah2012Completely,2011Achieving,2014On,Hao2016Achievable,6530617,6874856,7605466,6111234,2015Retrospective,2015Retrospective2,8600208,9145529,9217497,2021Two}. However, the information delay caused by IA technique has not been well studied yet. In order to quantify the information delay, a new criterion named degree of delay (DoD) is presented. It takes both queuing delay of transmitted messages \cite{shortle2018fundamentals} and the importance of the messages based on time-delay sensitivity \cite{536364} into account, which is vital for delay-sensitive networks. In this section, we first introduce the concept of DoD and analyze its influence factors. Then, we characterize the achievable DoD for three typical IA schemes, i.e., TDMA, BD-TDMA, and RIA.  Finally, we perform the performance investigation to prove the analysis mentioned above.

\subsection{Degree of Delay}
In interference networks, quality of service (QoS) \cite{536364} is taken to measure the performance of networks and its focus is mainly on two sides, i.e., on the user side and on the base station side. For the user side, two performance indicators, i.e., real-time and reliability, are crucial. However, recent studies of IA techniques consider more about reliability while real-time has been ignored. For the base station side, the base station provides various services, some of which are sensitive to the time delay. For these time-sensitive services, the base station is supposed to grant them higher priorities to guarantee their transmission first. On the other hand, for IA technique works, few depict the relationship between time-sensitive services and time delay. Thus, in view of the issues above, we first design the delay sensitive factor (DSF) to quantify the delay-tolerant ability of each service. Then, we associate DSF with time delay.

Without loss of generality, we consider that the base station provides services $\{1, \ldots , B\}$, which correspond to delay-tolerant levels in ascending order, i.e., $\{DTL1, \ldots ,DTLB\}$. According to the distribution of services, delay-tolerant levels are mapped to $P$ kinds of priority level \cite{KOCAREV2001199} and each of them is assigned delay sensitive factors, as shown in Table I.
\begin{table}[h]
\centering
\renewcommand\arraystretch{0.5}
  \centering
 \setlength{\abovecaptionskip}{0.cm}
 \setlength{\belowcaptionskip}{-0.2cm}
  \caption{mapping relation between priority and DSF }
{
	\begin{tabular}[b]{ccc}
   \cr \toprule[1.2pt]
delay-tolerant levels & priority level & delay sensitive factor
\cr \toprule[1.2pt]
\cr $DTL1,\ldots , DTLb_1$ & 1 & 1.0\\
  \cr $DTL(b_1\!+\!1),\ldots , DTLb_2$ & 2 & ${(P-1)}/P$\\
  \cr $\vdots$ & $\vdots$ & $\vdots$ \\
  \cr $DTL(b_{P-2}\!+\!1), \ldots , DTLb_{P-1}$ & ${P-1}$ & $2/P$\\
  \cr $DTL(b_{P-1}\!+\!1),\ldots , DTLB$ & $P$ & $1/P$\\
  \cr \bottomrule[1.2pt]
  \end{tabular}
}
  \label{tab:addlabel}
\end{table}

From Table I, we observe that the higher priority level to which the DTL corresponds, the more sensitive the information is to the time delay. Besides, the priority 1 represents the highest level among $P$ categories and its DSF value is the largest. It means that the information is the most sensitive to time delay and should be transmitted first. Based on these, the definition of DoD is derived. For ease of analysis, we set $P=5$ in this paper.

On the premise of reliable communication, we assume that the whole time elapsed in the communication process as total time delay ${D_s}$ \cite{robertazzi2000computer}. That is, the time period starts at the moment when the desired symbol $s$ is sent out, and lasts till the moment when the symbol is decoded. During this time, ${D_s}$ consists of waiting delay ${D_w}$, propagation delay ${D_p}$, and decoding delay ${D_{dc}}$, which is expressed by
\begin{equation}
\label{eq6}
{D_s} = {D_w} + {D_p} + {D_{dc}}.
\end{equation}
From (\ref{eq6}), if the positions of transmitter and receiver are fixed, the value of ${D_p}$ is constant. Meanwhile, ${D_{dc}}$ is the infinitesimal quantity of ${D_p}$, i.e., ${D_{dc}} = o({D_p})$. Therefore, ${D_s}$ can be regarded as a function of ${D_w}$. Taking the coherence time ${T_c}$ as the unit of time delay and the slots of waiting time ${n_s}$ as the quantity of time delay, the total time delay of desired symbol $s$ is calculated by
\begin{equation}
\label{eq7}
{D_s} = DS{F_s} \cdot {n_s} \cdot {T_c},
\end{equation}
where $DSF_s$ denotes the DSF of symobol $s$. Then, we define the data set ${\boldsymbol{S}_{num}}$ as the unit data set, and its size is $Num$. Accordingly, the system delay ${D_{sum}}$ is the weighted sum of the total delay ${D_s}$ where $s \in {\boldsymbol{S}_{num}}$, that is
\begin{equation}
\label{eq8}
{D_{sum}} = \sum\nolimits_{s \in {\boldsymbol{S}_{num}}}{D_s}.
\end{equation}
Herein, we introduce the DoD as follows:
\begin{equation}
\label{eq9}
DoD = \frac{{{D_{sum}}}}{{{T_c} \cdot Num}}= \frac{{\sum\nolimits_{s \in {\boldsymbol{S}_{num}}} {DS{F_s} \cdot {n_s}} }}{{Num}}.
\end{equation}
From (\ref{eq9}), DoD consists of three factors, which are delay sensitive factor $DSF$, queueing delay slot ${n_s}$, and size of data set ${Num}$. With further analysis, we find that DoD captures two performance indicators. Firstly, ${n_s}$ represents time cost of transmitting data set and its value depends on the capability of transmitting and receiving symbols within one slot. Then, the corresponding parameters are $M$ and $N$, respectively, and ${Num}$ can be viewed as normalization of the time cost. Secondly, $DSF$ reflects the QoS cost caused by time delay, whose value depends on the distribution of symbols with different priority levels. Herein, the issues about transmission delay and service classification can be considered with DoD.

\subsection{DoD of Typical Schemes}
We consider that the configuration of $K$-user MIMO BC network is $(K,M,N)$ where $M>N$, i.e., none of users can achieve decoding within one slot. Assume that a data set is sent from the base station and this set is constituted by $A$ symbols with priority $1$ and $B$ symbols with priority $5$. We take three kinds of typical schemes to transmit the data set, i.e., TDMA scheme, RIA scheme, and BD-TDMA scheme \cite{6414790}.

For TDMA scheme, $N$ symbols are sent from the base station side in each slot. To avoid DoF loss caused by round up operation $\left\lceil {\rm{*}} \right\rceil $, we assume that both $A$ and $B$ can be divided by $N$. Thus, $A/N$ slots and $B/N$ slots are required to send $A$ symbols with priority $1$ and $B$ symbols with priority $5$, respectively. Setting $A + B$ symbols as the size of the unit data set ${\boldsymbol{S}_{num}}$, the DoD of TDMA scheme is formulated as follows:
\begin{equation}
\label{eq10}
DoD_{TDMA} = \frac{{4{A^2}}}{{10(A + B)N}} + \frac{{A + B}}{{10N}} - \frac{{5A + B}}{{10(A + B)}}.
\end{equation}
Due to the restriction of $M>N$, TDMA scheme causes inefficient use of transmission, leading to an increase in the number of waiting slots ${n_s}$. Thus, its DoD is relatively high.

For BD-TDMA scheme, the communication process is the same as BD-TDMA scheme while $M$ symbols are sent during each slot. Thus, the DoD of BD-TDMA scheme is characterized by
\begin{equation}
\label{eq11}
DoD_{BD-TDMA} = \frac{{4{A^2}}}{{10(A + B)M}} + \frac{{A + B}}{{10M}} - \frac{{5A + B}}{{10(A + B)}}.
\end{equation}
Compared with TDMA scheme, for the sake of full utilization of space resources, the number of waiting slots ${n_s}$ gets decreased and the DoD of BD-TDMA scheme is relatively low. However, it adopts beamforming technique which incurs heavy computational burden to the transmitter.

For RIA scheme, the communication process is made of two stages, as shown in Fig. 3. 
\begin{figure}[h]
\vspace{-1.0em}
\centering
\includegraphics[width=6.5in]{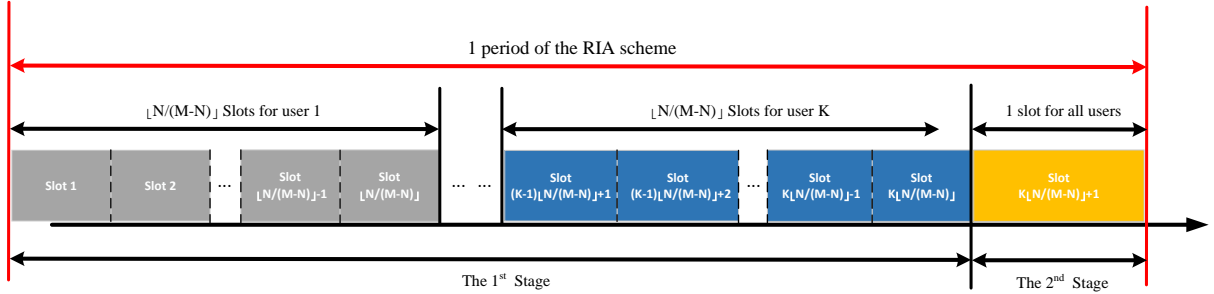}
\caption{one peroid of the RIA scheme}
\label{Fig3}
\end{figure}
For the first $\bar \varphi  = K\left\lfloor {{N \mathord{\left/{\vphantom {N {(M - N)}}} \right.\kern-\nulldelimiterspace} {(M - N)}}} \right\rfloor $ slots of first stage, the base station sends $M$ symbols in each slot. In seceond stage, the base station takes one slot to complete interference management of $K$ users simultaneously. We define $\bar \varphi +1 $ slots as the period of RIA scheme. Thus, the amount of $R = \left\lceil {{{(A + B)} \mathord{\left/{\vphantom {{(A + B)} {(\bar \varphi N + KN)}}} \right.\kern-\nulldelimiterspace} {(\bar \varphi N + KN)}}} \right\rceil $ periods is taken to transmit signals. Then, we further assume that both $A$ and $B$ are divisible by ${(\bar \varphi  + K)N}$. Otherwise, after $R$ periods of RIA scheme, one addtional period should be taken to transmit the rest of symbols, i.e., $A+B-R{(\bar \varphi N + KN)}$, which will cause a waste of time resources. Therefore, ${{[A(\bar \varphi  + 1)]} \mathord{\left/{\vphantom {{[A(\bar \varphi  + 1)]} {[(\bar \varphi  + K)N]}}} \right. \kern-\nulldelimiterspace} {[(\bar \varphi  + K)N]}}$ slots and ${{[B(\bar \varphi  + 1)]} \mathord{\left/
 {\vphantom {{[B(\bar \varphi  + 1)]} {[(\bar \varphi  + K)N]}}} \right.\kern-\nulldelimiterspace} {[(\bar \varphi  + K)N]}}$ slots are required to send $A$ symbols with priority $1$ and $B$ symbols with priority $5$, respectively. Setting $A + B$ symbols as the size of the unit data set ${\boldsymbol{S}_{num}}$, the DoD of RIA scheme is
\begin{equation}
\label{eq12}
\begin{split}
DoD_{RIA} = \frac{{(\bar \varphi  + 1)(A + B)}}{{10(\bar \varphi  + K)N}} + \frac{{4(\bar \varphi  + 1){A^2}}}{{10(\bar \varphi  + K)N(A + B)}} + \frac{{(\bar \varphi  - 1)(5A + B)}}{{10(A + B)}}.
\end{split}
\end{equation}

Hereto, the DoDs for the schemes are derived. To further analyze the DoD performance of these schemes, we next take simulations of DoD with variable $DSF$, ${n_s}$ and ${Num}$.

\subsection{Numerical Results of DoD}
In order to get a sense of the DoD definition and explore how does the DoD is affected by different system configurations, which will conduct our design of the IA schemes to obtain better DoD and DoF performane. In this subsection, some numerical rsults of DoD are given. Specifically, the configuration of the network is set as $(K,M,N) =(2,4,3)$. As mentioned before, DoD mainly depends on two aspects, i.e., the QoS cost and the time cost. To analyze how each factor does influence the value of DoD, we take simulations on two aspects separately.

Firstly, we analyze how the DoD is affected by $DSF$. To avoid the impact of non-integral period on DoD, we take total number of symbols as ${Num=2400}$, and the number of each type, i.e., symbols with priority $1$ and $5$, changes in multiples of $24$, e.g., $24$, $48$ and so forth. From Fig. 4, two points can be observed that, on the one hand, DoD has an exponential increasing relationship with the proportion of symbols with priority $1$. In specific, when the proportion of symbols with priority $1$ is 0\%, the DoDs of TDMA scheme, BD-TDMA scheme, and RIA scheme are 79.5, 59.5, and 70.5, respectively. When the proportion of symbols with priority $1$ accounts for 100\%, the DoDs of three schemes are 392.7, 294.3, and 346.9, respectively. The DoDs of the latter case, i.e., the proportion of symbols with priority 1 is 100\%, are 5 times greater than that of the former case, which is consistent with the change of DSF. On the one hand, whatever the proportion of symbols with priority $1$ is occupied with, the ranking order of DoDs under three schemes does not change. We also observe that the DoD of RIA scheme is 18\% higher than that of BD-TDMA scheme and meanwhile, the DoD of TDMA scheme exceeds the DoD of RIA scheme by 13\%. The reason is that RIA scheme makes better use of space resources than TDMA scheme, while it cannot decode the signals in real time like BD-TDMA scheme do. Thus, the DoD of RIA scheme is not the optimal among three schemes.
\begin{figure}[h]
\vspace{-1.0em}
\centering
\includegraphics[width=2.5in]{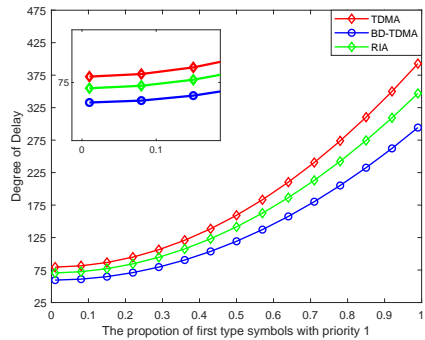}
\caption{The DoD affected by the QoS cost.}
\label{Fig4}
\end{figure}

Then, we analyze how is the DoD affected by ${n_s}$ and ${Num}$ under the condition that the proportion of symbols with priority $1$ and that of symbols with priority $5$ are fixed. In specific, we take the same amount of symbols from both types, and change the number of each type from 24 to 1200, which is in multiples of $24$. Then, we take the whole selected symbols as the unit data set ${\boldsymbol{S}_{num}}$ whose size is $Num=A+B$. From Fig. 5, it shows that the DoD affected by the time cost, i.e., ${n_s}$ increases with the growth of ${Num}$ and DoD increases synchronously. Meanwhile, the growth rate of DoD reflects the degrees that different transmission schemes are affected by ${n_s}$. In specific, when ${n_s}$ increases, the DoD of TDMA scheme grows fastest, followed by the DoD of RIA scheme, and the DoD of BD-TMDA scheme is least affected. This lies in fact that, on the one hand, TDMA scheme sends fewer symbols than the other two schemes within one slot, i.e., $M>N$. It means that more slots are taken to complete the task of sending data set, thereby, for each symbol, its number of waiting slots is the largest. On the other hand, the number of transmitted symbols of RIA scheme is the same as that of BD-TDMA scheme, while the former decodes all symbols of previous slots in the last slot, thereby its number of waiting slots is relatively large. Additionally, the selection of ${\boldsymbol{S}_{num}}$ affects the rationality of the simulation results. Specifically, when ${Num}$ is large, corresponding ${n_s}$ is so large that the information delay of each period has little impact on DoD of whole network. The same conclusion can be obtained from the figure that, when we take ${Num=48}$, the DoD of RIA scheme is 145.7\% larger than that of BD-TDMA scheme, and when we take ${Num=2400}$, the DoD of RIA scheme is 18.6\% larger than that of BD-TDMA scheme. In general, we take all of symbols within one period as ${Num}$ to make value of DoD more reasonable.
\begin{figure}[h]
\vspace{-1.0em}
\centering
\includegraphics[width=2.5in]{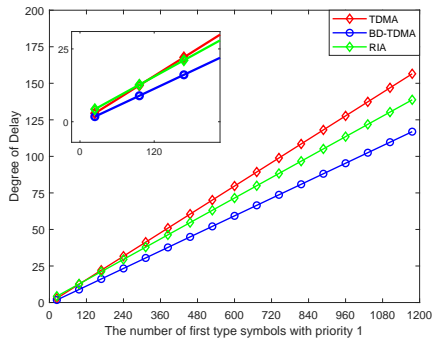}
\caption{The DoD affected by the time cost.}
\label{Fig5}
\end{figure}

After selecting appropriate data set size and determining time-delay sensitivity of the datas, DoD characterize the degree of information delay caused by transmission schems. Accordingly, we propose three novel schemes to improve DoD performance.  

\section{Hybrid Antenna Array Based Partial Interference Elimination and Retrospective Interference Regeneration Scheme}
As mentioned before, most of IA techniques, e.g., RIA schemes, complete decoding of all symbols in the last slot. This results in the fact that the received information cannot be acquired in time, since information delay occurs. Although we introduce DoD to quantify information delay, the specific method to solve this problem is still missing. Thus, we propose hybrid antenna array based partial interference elimination and retrospective interference regeneration scheme (HAA-PIE-RIR). The core idea of this scheme is adopting relay \cite{9200923} and designing a hybrid antenna array structure, i.e., combine omnidirectional antennas with directional antennas. From which, IUI can be eliminated and redundant symbols in desired signals are erased. Following that, distributed retrospective interference regeneration (DRIR) is presented to realize multiplexing of relay antennas. In this section, we first present HAA-PIE-RIR scheme where the configuration of the network is set as $\left( {K = 2,M,N} \right)$. Then, we introduce its improved scheme, i.e., hybrid antenna array based improved partial interference elimination and retrospective interference regeneration scheme (HAA-IPIE-RIR), to make it applicable in scenarios of $K$-user MIMO BC networks. Both DoF and DoD of these two schemes are then derived.

\subsection{HAA-PIE-RIR Scheme}
HAA-PIE-RIR scheme is mainly applied to $2$-user scenarios and the interference management process of the scheme can be divided into two phases, where the first phase spans $\varphi  = \left\lfloor {{N \mathord{\left/ {\vphantom {N {\left( {{\left\lceil {M/2} \right\rceil } - N} \right)}}} \right. \kern-\nulldelimiterspace} {\left( {{\left\lceil {M/2} \right\rceil } - N} \right)}}} \right\rfloor $ slots and the second phase takes $1$ slot. For each slot of the first phase $t \in \left\{ {1, \ldots ,\varphi } \right\}$, we assume that $2$-clusters of the base station transmit signals simultaneously, that is the cluster $1$ transmits symbols ${{\boldsymbol{S}}_1}[t] = \{ {s_{1}}[t], \ldots ,{s_{\left\lceil {M/2} \right\rceil }}[t]\} $ to the user $1$ with BC and the cluster $2$ transmits symbols ${{\boldsymbol{S}}_2}[t] = \{ {s_{\left\lceil {M/2} \right\rceil  + 1}}, \ldots ,{s_M}[t]\} $ to the user $2$ with BC. Then, the signals received by user $1$ and user $2$ are (\ref{eq13}) and (\ref{eq14}) respectively,
\begin{equation}
\label{eq13}
{{\boldsymbol{y}}_1}[t] = {\bf{H}}_1^1[t]{{\boldsymbol{S}}_1}[t] + {\bf{H}}_2^1[t]{{\boldsymbol{S}}_2}[t],
\end{equation}
\begin{equation}
\label{eq14}
{{\boldsymbol{y}}_2}[t] = {\bf{H}}_1^2[t]{{\boldsymbol{S}}_1}[t] + {\bf{H}}_2^2[t]{{\boldsymbol{S}}_2}[t],
\end{equation}
where ${\bf{H}}_i^j[t]$ denotes the channel matrix from cluster $i\in \{1,2\}$ to user $j\in \{1,2\}$. Under the condition of $M>N$, no user can decode desired symbols within one slot. To cope with this issue, the components of the received signals should be studied. By analyzing (\ref{eq13}) and (\ref{eq14}), we find that the received signals are the combinations of desired signals and IUI. Thus, we mark desired signals as ${\boldsymbol{L}}_i^j[t]$ and IUI as ${\boldsymbol{I}}_i^j[t]$. Then, the form of received signals is changed to
\begin{equation}
\label{eq15}
{{\boldsymbol{y}}_1}[t] = {\boldsymbol{L}}_1^1[t] + {\boldsymbol{I}}_2^1[t],
\end{equation}
\begin{equation}
\label{eq16}
{{\boldsymbol{y}}_2}[t] = {\boldsymbol{I}}_1^2[t] + {\boldsymbol{L}}_2^2[t].
\end{equation}
To eliminate IUI, the relay technique is adopted and its structure is modified to the hybrid antenna array. Regarding the relay, we suppose that three preconditions hold.
\begin{itemize}
\item When the clusters of base station transmit signals to users, relay $R$ can also receive signals from base station.
\item The number of relay antennas $Q$ is adequate, i.e., $Q \ge \max {{\{ M, \max\{ }}\left\lceil {M/2} \right\rceil {{,N\}  + 2N\} }}$.
\item Relay $R$ can acquire moderately delayed CSI and delayed CSI of the whole network.
\end{itemize}

Under the above conditions, the symbols sent by the base station can be captured by the relay. Specifically, when relay $R$ receives signals, that is
\begin{equation}
\label{eq17}
{{\boldsymbol{y}}_R}[t] = {\bf{H}}_B^R[t]\{{\boldsymbol{S}}_1[t],{\boldsymbol{S}_2}[t]\},
\end{equation}
where ${\bf{H}}_B^R[t]$\footnote{It is known that imperfect CSI is always an important problem for the IA. Generally, the issue of imperfect CSI can be summarized into three cases, i.e., the accurate and delayed CSI, the inaccurate and instaneous CSI,and the inaccurate and delayed CSI. In this paper, our focus is on DoF and DoD, which corresponds to the case of accurate and delayed CSI. How to extent the study to the other two cases is an interesting problem but it is not in the scope of this paper.} denotes channel matrix from the base station to relay $R$. From (\ref{eq17}), all symbols can be decoded, i.e., $\{{\boldsymbol{S}}_1[t],{\boldsymbol{S}_2}[t]\}$. Meanwhile, the moderately delayed CSIs are fed to relay $R$, where both ${\bf{H}}_R^1{[t]}$ and ${\bf{H}}_R^2{[t]}$ are known at relay side. Relay $R$ designs transmitted signals as
\begin{equation}
\label{eq18}
{{\boldsymbol{x}}_R}[t] =  - ({\bf{H}}_R^1{[t]^{ - 1}}{\boldsymbol{I}}_2^1[t] + {\bf{H}}_R^2{[t]^{ - 1}}{\boldsymbol{I}}_1^2[t]),
\end{equation}
where ${\bf{H}}_R^j{[t]},j \in \{1,2\}$ denotes channel matrix from relay $R$ to user $j$. Note that ${\boldsymbol{I}}_2^1[t]$ and ${\boldsymbol{I}}_1^2[t]$ are $N  \times  {\left\lceil {M/2} \right\rceil }$ dimensional matrices, and therefore the number of ${{\max\{ }}\left\lceil {M/2} \right\rceil {{,N\} }}$ relay antennas is required to transmit signals. After receiving signals from relay $R$, received signals of users $1$ and $2$ are presented in (\ref{eq19}) and (\ref{eq20}) respectively,
\begin{small}
\begin{equation}
\label{eq19}
\begin{split}
{\boldsymbol{{\bar y}}_1}[t] &= {\boldsymbol{L}}_1^1[t] + {\boldsymbol{I}}_2^1[t] +{\bf{H}}_R^1{[t]}{\boldsymbol{x}}_R{[t]}\\
 &= ({\bf{H}}_1^1[t] - {\bf{H}}_R^1[t]{\bf{H}}_R^2{[t]^{ - 1}}{\bf{H}}_1^2[t]){{\boldsymbol{S}}_1}[t]\\
 &={{{\bf{\bar H}}}^1_1}[t]{{\boldsymbol{S}}_1}[t],
\end{split}
\end{equation}
\begin{equation}
\label{eq20}
\begin{split}
{\boldsymbol{{\bar y}}_2}[t] &= {\boldsymbol{I}}_1^2[t] + {\boldsymbol{L}}_2^2[t] +{\bf{H}}_R^2{[t]}{\boldsymbol{x}}_R{[t]}\\
 &=({\bf{H}}_2^2[t] - {\bf{H}}_R^2[t]{\bf{H}}_R^1{[t]^{ - 1}}{\bf{H}}_2^1[t]){{\boldsymbol{S}}_2}[t]\\
 &={{{\bf{\bar H}}}^2_2}[t]{{\boldsymbol{S}}_2}[t],
\end{split}
\end{equation}
\end{small}
where ${{{\bf{\bar H}}}^1_1}[t]$ and ${{{\bf{\bar H}}}^2_2}[t]$ denote equivalent channel matrices to user $1$ and user $2$, respectively. With the help of relay, IUI gets eliminated. However, the relationship between $\left\lceil {M/2} \right\rceil$ and $N$ needs to be further discussed. Specifically, if $\left\lceil {M/2} \right\rceil {{ \le  N}}$, the homogeneous solutions of matrices ${{{\bf{\bar H}}}^1_1}[t]$ and ${{{\bf{\bar H}}}^2_2}[t]$ exist. Otherwise, the matrices are still underdetermined. For the former case, all the symbols can be decoded within one slot so that the second phase of HAA-PIE-RIR scheme is unnecessary. For the latter case, we reconstruct the structure of antennas array belonging to relay $R$ and send partial interference signals with directional antennas, where redundant symbols in desired signals get erased.

In specific, according to moderately delayed CSI, the ranks of channel matrices, i.e., ${{{\bf{\bar H}}}^1_1}[t]$ and ${{{\bf{\bar H}}}^2_2}[t]$, are determined. It means that the number of users' antennas can be confirmed, i.e., $rank({{{\bf{\bar H}}}^1_1}[t]) = N$ and $rank({{{\bf{\bar H}}}^2_2}[t]) = N$. Then, relay $R$ takes every $N$ additional directional antennas to serve each of users, respectively. From which, partial interference elimination signals of user $1$ and user $2$, i.e., ${{\boldsymbol{\hat x}}_R^1[t]}$ and ${{\boldsymbol{\hat x}}_R^2[t]}$, are designed as follows
\begin{equation}
\label{eq21}
\begin{small}
\setlength{\arraycolsep}{0.5pt}
\begin{array}{c}
\underbrace {\left[ {\begin{array}{*{20}{c}}
{{x_{\left\lceil {M/2} \right\rceil  + 1}}[t]}\\
 \vdots \\
{{x_{\left\lceil {M/2} \right\rceil  + N}}[t]}
\end{array}} \right]}_{{\boldsymbol{\hat x}}_R^1[t]}{{ = }}{\underbrace {\left[ {\begin{array}{*{20}{c}}
{h_{R,\left\lceil {M/2} \right\rceil  + 1}^{1,1}[t]}& \cdots &{h_{R,\left\lceil {M/2} \right\rceil  + N}^{1,1}[t]}\\
 \vdots & \ddots & \vdots \\
{h_{R,\left\lceil {M/2} \right\rceil  + 1}^{1,N}[t]}& \cdots &{h_{R,\left\lceil {M/2} \right\rceil  + N}^{1,N}[t]}
\end{array}} \right]}_{{\bf{\hat H}}_R^1[t]}}^{ - 1} \times \left[ {\begin{array}{*{20}{c}}
{\sum\limits_{m = N + 1}^{\left\lceil {M/2} \right\rceil } { - \bar h_{1,m}^{1,1}[t]{s_m}}[t] }\\
 \vdots \\
{\sum\limits_{m = N + 1}^{\left\lceil {M/2} \right\rceil } { - \bar h_{1,m}^{1,N}[t]{s_m}}[t] }
\end{array}} \right],
\end{array}
\end{small}
\end{equation}
\begin{equation}
\label{eq22}
\begin{small}
\setlength{\arraycolsep}{0.5pt}
\begin{array}{c}
\underbrace {\left[ {\begin{array}{*{20}{c}}
{{x_{\left\lceil {M/2} \right\rceil  + N + 1}}[t]}\\
 \vdots \\
{{x_{\left\lceil {M/2} \right\rceil  + 2N}}[t]}
\end{array}} \right]}_{{\boldsymbol{\hat x}}_R^2[t]}{{ = }}{\underbrace {\left[ {\begin{array}{*{20}{c}}
{h_{R,\left\lceil {M/2} \right\rceil  + N + 1}^{2,1}[t]}& \cdots &{h_{R,\left\lceil {M/2} \right\rceil  + 2N}^{2,1}[t]}\\
 \vdots & \ddots & \vdots \\
{h_{R,\left\lceil {M/2} \right\rceil  + N + 1}^{2,N}[t]}& \cdots &{h_{R,\left\lceil {M/2} \right\rceil  + 2N}^{2,N}[t]}
\end{array}} \right]}_{{\bf{\hat H}}_R^2[t]}}^{ - 1} \times \left[ {\begin{array}{*{20}{c}}
{\sum\limits_{m = \left\lceil {M/2} \right\rceil {{ + }}N + 1}^M { - \bar h_{2,m}^{2,1}[t]{s_m}}[t] }\\
 \vdots \\
{\sum\limits_{m = \left\lceil {M/2} \right\rceil {\rm{ + }}N + 1}^M { - \bar h_{2,m}^{2,N}[t]{s_m}}[t] }
\end{array}} \right],
\end{array}
\end{small}
\end{equation}
where ${{\bf{\hat H}}_R^1[t]}$ and ${{\bf{\hat H}}_R^2[t]}$ are the channel matrices between relay $R$'s directional antennas and users' antennas. After receiving signals from directional antennas of relay $R$, received signals of users $1$ and $2$ are converted to (\ref{eq23}) and (\ref{eq24}) respectively,
\begin{small}
\begin{equation}
\label{eq23}
\setlength{\arraycolsep}{0.5pt}
\begin{split}
{{{\boldsymbol{\hat y}}}_1}[t] &= {{{\boldsymbol{\bar y}}}_1}[t] + {\bf{\hat H}}_{R}^{1}[t]{\boldsymbol{\hat x}}_{R}^{1}{{[t]}}
= \left[ {\begin{array}{*{20}{c}}
{\bar h_{1,1}^{1,1}[t]}& \cdots &{\bar h_{1,N}^{1,1}[t]}\\
 \vdots & \ddots & \vdots \\
{\bar h_{1,1}^{1,N}[t]}& \cdots &{\bar h_{1,N}^{1,N}[t]}
\end{array}} \right]\left[ {\begin{array}{*{20}{c}}
{{s_1}[t]}\\
 \vdots \\
{{s_N}[t]}
\end{array}} \right],
\end{split}
\end{equation}
\begin{equation}
\label{eq24}
\setlength{\arraycolsep}{0.5pt}
{{{\boldsymbol{\hat y}}}_2}[t] = {{{\boldsymbol{\bar y}}}_2}[t] + {\bf{\hat H}}_R^2[t]{\boldsymbol{\hat x}}_R^2[t]=\left[ {\begin{array}{*{20}{c}}
{\bar h_{2,\left\lceil {M/2} \right\rceil  + 1}^{2,1}[t]}& \cdots &{\bar h_{2,\left\lceil {M/2} \right\rceil  + N}^{2,1}[t]}\\
 \vdots & \ddots & \vdots \\
{\bar h_{2,\left\lceil {M/2} \right\rceil  + 1}^{2,N}[t]}& \cdots &{\bar h_{2,\left\lceil {M/2} \right\rceil {{ + N}}}^{2,N}[t]}
\end{array}} \right]\left[ {\begin{array}{*{20}{c}}
{{s_{\left\lceil {M/2} \right\rceil  + 1}}[t]}\\
 \vdots \\
{{s_{\left\lceil {M/2} \right\rceil  + N}}[t]}
\end{array}} \right].
\end{equation}
\end{small}
Herein, the space of received signals is reduced to a lower dimension where the desired symbols can be decoded within one slot. Although utilizing relay enables all users to decode desired symbols in every slot, the redundant symbols that have been eliminated are also indispensable for users. Thus, in the second phase, DRIR algorithm is adopted to compensate symbols which are missing in previous slots.

In the second phase, only one slot, i.e., $t=\varphi + 1$, is taken to achieve DRIR algorithm. In specific, with combined CSI and the knowledge of missing symbols, relay $R$ regenerates the eliminated part of desired symbols as
\begin{equation}
\label{eq25}
\setlength{\arraycolsep}{0.5pt}
\begin{array}{l}
{{{\boldsymbol{\hat x}}}_{_{{R}}}}{\rm{[}}\varphi {\rm{ + 1] = }}\left[ {\begin{array}{*{20}{c}}
{{\boldsymbol{\hat x}}_R^1{\rm{[}}\varphi {\rm{ + 1]}}}\\
{{\boldsymbol{\hat x}}_R^2{\rm{[}}\varphi {\rm{ + 1]}}}
\end{array}} \right]{\rm{ = }}\left[ {\begin{array}{*{20}{c}}
{{\bf{\hat H}}_R^1{{[\varphi {\rm{ + 1}}]}^{ - 1}}}&{{{\bf{0}}_{N \times N}}}\\
{{{\bf{0}}_{N \times N}}}&{{\bf{\hat H}}_R^2{{[\varphi {\rm{ + 1}}]}^{ - 1}}}
\end{array}} \right] \times \\
\begin{array}{l}
[{s_{N + 1}}[1], \ldots ,{s_{\left\lceil {M/2} \right\rceil }}[1],{s_{N + 1}}[2], \ldots ,{s_{\left\lceil {M/2} \right\rceil }}[\varphi ],\\
{s_{\left\lceil {M/2} \right\rceil {\rm{ + }}N + 1}}[1], \ldots ,{s_M}[1],{s_{\left\lceil {M/2} \right\rceil {\rm{ + }}N + 1}}[2], \ldots ,{s_M}[\varphi ]]_.^T
\end{array}
\end{array}
\end{equation}
Due to the beamforming nature of directional antennas, both user $1$ and user $2$ receive its missing signals without IUI. The received signals are expressed by
\begin{equation}
\label{eq26}
{{\boldsymbol{\hat y}}_1}{\rm{[}}\varphi {\rm{ + 1] = }}{\bf{\hat H}}_R^1[\varphi {\rm{ + 1}}]{\boldsymbol{\hat x}}_R^1{\rm{[}}\varphi {\rm{ + 1]}},
\end{equation}
\begin{equation}
\label{eq27}
{{\boldsymbol{\hat y}}_2}{\rm{[}}\varphi {\rm{ + 1] = }}{\bf{\hat H}}_R^2[\varphi {\rm{ + 1}}]{\boldsymbol{\hat x}}_R^2{\rm{[}}\varphi {\rm{ + 1]}}.
\end{equation}
From (\ref{eq26}), the missing symbols ${\rm{\{ }}{s_{N + 1}}[1], \ldots ,{s_{\left\lceil {M/2} \right\rceil }}[1]$ $,{s_{N + 1}}[2], \ldots ,{s_{\left\lceil {M/2} \right\rceil }}[\varphi ]{\rm{\} }}$ can be gotten by user $1$ and, similar to (\ref{eq27}), $\{ {s_{\left\lceil {M/2} \right\rceil {\rm{ + }}N + 1}}[1], \ldots ,{s_M}[1], {s_{\left\lceil {M/2} \right\rceil {\rm{ + }}N + 1}}[2], $ $\ldots ,{s_M}[\varphi ]\} $ can be obtained by user $2$ simultaneously.

\subsection{HAA-IPIE-RIR Scheme}
For $2$-user $M \times N$ BC networks, HAA-PIE-RIR scheme realizes decoding in each slot, and hence the problem of information delay can be relieved. Unfortunately, when we attempt extending HAA-PIE-RIR scheme to the application scenarios with $K$-user, it is hard to the decrease dimension of interference space into an acceptable value, where IUI can be eliminated thoroughly. Therefore, we further propose its improved scheme, i.e., HAA-IPIE-RIR scheme. The improvement idea of this scheme is that, in the first phase, the relay $R$ first makes a combination process among interference signals from different clusters, and then designs the precoding matrix with the combination results. This achieves the purpose of eliminating IUI of each user simultaneously.

Specifically, the first phase spans $\varphi  = \left\lfloor {{N \mathord{\left/ {\vphantom {N {\left( {{\left\lceil {M/K} \right\rceil } - N} \right)}}} \right. \kern-\nulldelimiterspace} {\left( {{\left\lceil {M/K} \right\rceil } - N} \right)}}} \right\rfloor $ slots and the second phase takes $1$ slot. For each slot of the first phase $t \in \left\{ {1, \ldots ,\varphi } \right\}$, clusters of the base station transmit signals $\{{{\boldsymbol{S}}_1}[t], \ldots ,{\boldsymbol{S}}_K[t]\}$, respectively. For the user $j,\forall j \in {\boldsymbol K}$, its received signals is expressed by
\begin{equation}
\label{eq28}
\begin{split}
{{\boldsymbol{y}}_j}[t] & = {\bf{H}}_j^j[t]{{\boldsymbol{S}}_j}[t] + \sum\limits_{i = 1\backslash j}^K {{\bf{H}}_i^j[t]{{\boldsymbol{S}}_i}[t} ]\\
& = {\boldsymbol{L}}_j^j[t] + \sum\limits_{i = 1\backslash j}^K {{\boldsymbol{I}}_i^j[t]} ,i \in \boldsymbol{I},j \in \boldsymbol{K},
\end{split}
\end{equation}
where ${\boldsymbol{L}}_j^j[t]$ is desired signal from cluster $j$ to user $j$ and meanwhile, ${\boldsymbol{I}}_i^j[t]$ denotes interference from cluster $i$ to user $j$. Suppose the three preconditions mentioned above still hold. Then, relay $R$ designs transmitted signals as
\begin{equation}
\label{eq29}
\begin{small}
\setlength{\arraycolsep}{0.5pt}
\begin{array}{c}
{{\boldsymbol{x}}_R}[t] =  -{\underbrace {\left[ {\begin{array}{*{20}{c}}
{{\bf{H}}_R^1[t]}\\
 \vdots \\
{{\bf{H}}_R^j[t]}\\
 \vdots \\
{{\bf{H}}_R^K[t]}
\end{array}} \right]}_{{{\bf{H}}_R}[t]}}^{ - 1}
 \!\!\!\!\!\!\!\!\times \left[ {\begin{array}{*{20}{c}}
{\bf{0}}& \cdots &{{\bf{H}}_i^1[t]}& \cdots &{{\bf{H}}_K^1[t]}\\
 \vdots & \ddots & \vdots & \ddots & \vdots \\
{{\bf{H}}_1^j[t]}& \cdots &{\bf{0}}& \cdots &{{\bf{H}}_K^j[t]}\\
 \vdots & \ddots & \vdots & \ddots & \vdots \\
{{\bf{H}}_1^K[t]}& \cdots &{{\bf{H}}_i^K[t]}& \cdots &{\bf{0}}
\end{array}} \right]\left[ {\begin{array}{*{20}{c}}
{{{\boldsymbol{S}}_1}}\\
 \vdots \\
{{{\boldsymbol{S}}_j}}\\
 \vdots \\
{{{\boldsymbol{S}}_K}}
\end{array}} \right],
\end{array}
\end{small}
\end{equation}
where ${\bf{H}}_R^j[t] \in {\boldsymbol{C}}^{N \times KN}$ denotes channel matrix from relay $R$ to user $j$ and ${{\bf{H}}_R}[t]$ is used to mark channel matrix from relay $R$ to all users. After receiving signals from relay $R$, the received signals of user $j$ is converted to
\begin{equation}
\label{eq30}
\setlength{\arraycolsep}{0.5pt}
{{\bf{y}}_j}[t] = {\bf{H}}_j^j[t]{{\boldsymbol{S}}_j}[t] + \sum\limits_{i = 1\backslash j}^K {{\bf{H}}_i^j[t]{{\boldsymbol{S}}_i}[t} ]+{\bf{H}}_R[t]{{\boldsymbol{x}}_R}[t]
 = {\boldsymbol{L}}_j^j[t].
\end{equation}
Herein, all IUIs from other users $- j = {\boldsymbol K}\backslash j$ are eliminated. Furthermore, the subsequent process of HAA-IPIE-RIR scheme is similar to HAA-PIE-RIR scheme, i.e., if $\left\lceil {M/K} \right\rceil  {{ \le N}}$, the second phase can be omitted. On the contrary, if $\left\lceil {M/K} \right\rceil {{ > N}}$, DRIR algorithm is adopted with relay $R$'s directional antennas.

\subsection{DoD and DoF of Two Schemes}
In this subsection, we analyze the DoD and DoF for above proposed two schemes in $K$-user $M \times N$ BC networks. We assume that the configuration is set to $(K,M,N)$ and the data set consists of $A$ symbols with priority $1$ and $B$ symbols with priority $5$. For this application scenario, we analyze DoD and DoF of proposed schemes, separately.

For HAA-PIE-RIR scheme, the number of slots that are used to transmit data set should be studied. In specific, if $\left\lceil {M/2} \right\rceil {{ \ >  N}}$, $\varphi  = \left\lfloor {{N \mathord{\left/ {\vphantom {N {\left( {{\left\lceil {M/2} \right\rceil } - N} \right)}}} \right. \kern-\nulldelimiterspace} {\left( {{\left\lceil {M/2} \right\rceil } - N} \right)}}} \right\rfloor $ slots and $1$ slot are taken in the first phase and the second phase, respectively. For this case, $2N$ symbols are decoded at users side within each slot of two phases. If $\left\lceil {M/2} \right\rceil {{ \le  N}}$, there is only the first phase and it also takes $\varphi$ incoherent slots. For this case, $M$ symbols are decoded at users side within each slot. By analyzing such cases, we find that the number of decoded symbols in average is constant. Thus, we merge two cases together, where $\min {\rm{\{ }}2N,M{\rm{\} }}$ symbols are  decoded at users side within one slot. Meanwhile, $ \varphi +1 $ slots can be regarded as the period of HAA-PIE-RIR scheme where $\min {\rm{\{ }}2N,M{\rm{\} }}(\varphi +1)$ symbols are obtained by users, and the total number of $R = \left\lceil {{{(A + B)} \mathord{\left/{\vphantom {{(A + B)} {[\min {\rm{\{ }}2N,M{\rm{\} }} \times (\varphi +1)}]}} \right. \kern-\nulldelimiterspace}[ \min {\rm{\{ }}2N,M{\rm{\} }}(\varphi +1)}]} \right\rceil $ periods are required to complete the transmission of symbols $A+B$. Note that both $A$ and $B$ should be divisible by $\min {\rm{\{ }}2N,M{\rm{\} }}(\varphi +1)$, otherwise one more period is taken to transmit remaining symbols, i.e., $A + B-R\min {\rm{\{ }}2N,M{\rm{\} }}(\varphi +1)$. From which, it leads to a waste of time resources. Besides, we assume that the transmission is reliable and the size of the unit data set ${\boldsymbol{S}_{num}}$ is $A + B$. The DoF and DoD of HAA-PIE-RIR scheme are given by (\ref{eq31}) and (\ref{eq32}) respectively,
\begin{equation}
\label{eq31}
DoF_{HAA-PIE-RIR} = \frac{{\min \{ 2N,M\} (\varphi  + 1)}}{{(\varphi  + 1)}} = \min \{ 2N,M\},
\end{equation}
\begin{equation}
\label{eq32}
DoD_{HAA-PIE-RIR}  = \frac{{(A + B)}}{{10\min \{ 2N,M\} }} + \frac{{4{A^2}}}{{10(A + B)\min \{ 2N,M\} }} - \frac{{5A + B}}{{10(A + B)}}.
\end{equation}

For HAA-IPIE-RIR scheme, the core idea of dealing with IUI and redundant symbols is consistent with HAA-PIE-RIR scheme. The main difference between the two algorithms is that relay $R$ changes its structure of precoding matrix where multiple users' IUIs are eliminated. Accordingly, the judgment threshold for different situations of the first phase becomes into $KN=M$ and meanwhile, the numbers of slots and obtained symbols in each period are changed into $\varphi  = \left\lfloor {{N \mathord{\left/ {\vphantom {N {\left( {{\left\lceil {M/K} \right\rceil } - N} \right)}}} \right. \kern-\nulldelimiterspace} {\left( {{\left\lceil {M/K} \right\rceil } - N} \right)}}} \right\rfloor $ and $\min \{ KN,M\} (\varphi  + 1)$, respectively. Herein, the DoF and DoD of HAA-PIE-RIR scheme are given by (\ref{eq33}) and (\ref{eq34}) respectively,
\begin{equation}
\label{eq33}
DoF_{HAA-IPIE-RIR} = \frac{{\min \{ KN,M\} (\varphi  + 1)}}{{(\varphi  + 1)}} = \min \{ KN,M\} ,
\end{equation}
\begin{equation}
\label{eq34}
DoD_{HAA-IPIE-RIR} =\frac{{(A + B)}}{{10\min \{ KN,M\} }} \! + \! \frac{{4{A^2}}}{{10(A \!+\! B)\min \{ KN,M\} }}- \frac{{5A + B}}{{10(A + B)}}.
\end{equation}
Comparing HAA-IPIE-RIR scheme with HAA-PIE-RIR scheme, when $M/N>2$,  the former achieves higher DoF and lower DoD. However, if the values of $M$ and $N$ are large, the computational burden of relay is heavy due to the precoding matrix design in the first phase.  The issue of computational cost is further discussed in Section VI.

\section{Hybrid Antenna Array Based Cyclic Interference Elimination and Retrospective Interference Regeneration Scheme }
For multi-user $M \times N$ BC networks, HAA-PIE-RIR scheme effectively improves both DoD and DoF in $2$-user application scenario. Subsequently, its improved scheme, i.e., HAA-IPIE-RIR scheme, is proposed and the application scenario gets extended to $K$-user. However, the scheme also brings considerable computational burden to the relay, which makes it not practically applicable. Therefore, two new techniques are considered. One is to introduce multiple relays \cite{8680727} which makes the space of IUI into a lower dimension. Although this way can eliminate IUIs of all users thoroughly, the deployment of relays becomes much more complex and interference among relays should be considered. The other is to take a trade-off tactic \cite{8340759} between space resources and time resources. Based on this idea, we propose hybrid antenna array based cyclic interference elimination and retrospective interference regeneration scheme (HAA-CIE-RIR). The core idea of this scheme is adopting a slot-based cyclic structure in the first phase, where $K$-user application scenario is simplified to multiple $2$-user application scenarios. Meanwhile, in the second phase, DRIR algorithm is taken to obtain multiplexing gain of $K$-user. In this section, we present HAA-CIE-RIR scheme under the condition of $\left( {K,M,N} \right)$ configuration and derive its DoF and DoD of the proposed scheme.

\subsection{HAA-CIE-RIR Scheme}
The HAA-CIE-RIR scheme consists of two phases where the first phase takes  $\left\lfloor {{N \mathord{\left/
 {\vphantom {N {(M - 2N)}}} \right.
 \kern-\nulldelimiterspace} {(M - 2N)}}} \right\rfloor $ groups of $K$ slots and the second phase spans $1$ slot. For each group of the first phase, the base station takes turns to provide services for $2$-users simultaneously. Specifically, for slot $j \in \left\{ {1, \ldots ,K } \right\}$ of $t$-th group where $t \in \{ 1, \ldots ,\left\lfloor {{N \mathord{\left/
 {\vphantom {N {(M - 2N)}}} \right.
 \kern-\nulldelimiterspace} {(M - 2N)}}} \right\rfloor\} $, the base station transmits signals ${{\boldsymbol{S}}_1}[j] = \{ {s_{1}}[j], \ldots ,{s_{\left\lceil {M/2} \right\rceil }}[j]\} $ to user $j$ with BC and ${{\boldsymbol{S}}_{2}}[j] = \{ {s_{{\left\lceil {M/2} \right\rceil }+1}}[j], \ldots ,{s_M}[j]\} $ to user $j+1$ with BC, respectively. Note that, when $j=K$, the first cluster transmits ${{\boldsymbol{S}}_1}[j] = \{ {s_{1}}[j], \ldots ,{s_{\left\lceil {M/2} \right\rceil }}[j]\} $ to user $\varphi$ and the second cluster transmits ${{\boldsymbol{S}}_{2}}[j] = \{ {s_{{\left\lceil {M/2} \right\rceil }+1}}[j], \ldots ,{s_M}[j]\} $ to user $1$. We denote all slots of the first phase as $\varphi $ where $\varphi=K\left\lfloor {{N \mathord{\left/
 {\vphantom {N {(M - 2N)}}} \right.
 \kern-\nulldelimiterspace} {(M - 2N)}}} \right\rfloor$. By analysing signals of the first phase, the distribution of desired symbols is uniform, i.e., each user receives $M$ desired symbols within two slots. Therefore, the proposed hybrid antenna array structure of relay $R$ is utilized.

Specifically, for slot $j \in \left\{ {1, \ldots ,K } \right\}$, the signals received by user $j$ and user $j+1$ are (\ref{eq35}) and (\ref{eq36}) respectively,
\begin{equation}
\label{eq35}
\begin{split}
{{\bf{y}}_j}[j] &= {\bf{H}}_1^j[j]{{\boldsymbol{S}}_1}[j] + {\bf{H}}_2^j[j]{{\boldsymbol{S}}_2}[j]\\
&= {\boldsymbol{L}}_1^j[j] + {\boldsymbol{I}}_2^j[j]
\end{split},
\end{equation}
\begin{equation}
\label{eq36}
\begin{split}
{{\bf{y}}_{j + 1}}[j] &= {\bf{H}}_1^{j + 1}[j]{{\boldsymbol{S}}_1}[j] + {\bf{H}}_2^{j + 1}[j]{{\boldsymbol{S}}_2}[j]\\
&= {\boldsymbol{I}}_1^{j{\rm{ + }}1}[j] + {\boldsymbol{L}}_2^{j{\rm{ + }}1}[j],
\end{split}
\end{equation}
where ${\bf{H}}_i^j[j]$ denotes the channel matrix from cluster $i\in \{1,2\}$ to user $j$ and meanwhile, ${\boldsymbol{L}}_i^j[j]$ and ${\boldsymbol{I}}_i^j[j]$ denote corresponding desired signals and interference, respectively. Suppose the three preconditions mentioned above still hold. Then, in slot $j$, the transmitted signals designed by relay $R$ are expressed by
\begin{equation}
\label{eq37}
{{\boldsymbol{x}}_R}[j] =  - ({\bf{H}}_R^j{[j]^{ - 1}}{\boldsymbol{I}}_2^j[j] + {\bf{H}}_R^{j+1}{[j]^{ - 1}}{\boldsymbol{I}}_1^{j+1}[j]).
\end{equation}
When users $j$ and $j+1$ receive signals from relay $R$ in slot $j$, the received signals are turned into (\ref{eq38}) and (\ref{eq39}) respectively,
\begin{equation}
\label{eq38}
\begin{split}
{{\bf{y}}_j}[j] &= {\bf{H}}_1^j[j]{{\bf{S}}_1}[j] + {\bf{H}}_2^j[j]{{\bf{S}}_2}[j]+{\bf{H}}_R^j{[j]}{\boldsymbol{x}}_R{[j]}\\
&= ({\bf{H}}_1^j[j] - {\bf{H}}_R^j[j]{\bf{H}}_R^{j + 1}{[j]^{ - 1}}{\bf{H}}_1^{j + 1}[j]{\rm{)}}{{\bf{S}}_1}[j]\\
&={{{\bf{\bar H}}}^j_1}[j]{{\bf{S}}_1}[j],
\end{split}
\end{equation}
\begin{equation}
\label{eq39}
\begin{split}
{{\boldsymbol{y}}_{j{+}1}}[j] &= {\bf{H}}_1^{j{ + }1}[j]{{\boldsymbol{S}}_1}[j] \!+\! {\bf{H}}_2^{j{ + }1}[j]{{\boldsymbol{S}}_2}[j] \!+\! {\bf{H}}_R^{j{ + }1}[j]{{\boldsymbol{x}}_R}[j]\\
&= ({\bf{H}}_2^{j{ + }1}[j] - {\bf{H}}_R^{j + 1}[j]{\bf{H}}_R^j{[j]^{ - 1}}{\bf{H}}_2^j[j]){{\boldsymbol{S}}_2}[j]\\
&={{{\bf{\bar H}}}^{j + 1}_2}[j]{{\boldsymbol{S}}_2}[j],
\end{split}
\end{equation}
where ${{{\bf{\bar H}}}^{j}_1}[j]$ and ${{{\bf{\bar H}}}^{j+1}_2}[j]$ represent equivalent channel matrices of user $j$ and user $j+1$, respectively. After eliminating IUIs of all users, the relationship between $\left\lceil {M/2} \right\rceil$ and $N$ should be further discussed. Specifically, if $\left\lceil {M/2} \right\rceil {{ \le  N}}$, the second phase is canceled. Otherwise, the hybrid antenna array structure of relay $R$ is adopted and DRIR algorithm is taken.

In specific, according to the ranks of matrices ${{{\bf{\bar H}}}^{j}_1}[j]$ and ${{{\bf{\bar H}}}^{j+1}_2}[j]$, the number of redundant symbols are determined, i.e., $\left\lceil {M/2} \right\rceil-rank({{{\bf{\bar H}}}^j_1}[j])$ and $M-\left\lceil {M/2} \right\rceil-rank({{{\bf{\bar H}}}^{j+1}_2}[j])$. With moderately delayed CSI, relay $R$ designs partial interference elimination signals of user $j$ and user $j+1$ respectively,
\begin{equation}
\label{eq40}
\begin{small}
\setlength{\arraycolsep}{0.5pt}
\begin{array}{*{20}{c}}
{\underbrace {\left[ {\begin{array}{*{20}{c}}
{{x_{\left\lceil {M/2} \right\rceil  + 1}}[j]}\\
 \vdots \\
{{x_{\left\lceil {M/2} \right\rceil  + N}}[j]}
\end{array}} \right]}_{{\bf{\hat x}}_R^j[j]}{\rm{ = }}{{\underbrace {\left[ {\begin{array}{*{20}{c}}
{h_{R,\left\lceil {M/2} \right\rceil  + 1}^{j,1}[j]}& \cdots &{h_{R,\left\lceil {M/2} \right\rceil  + N}^{j,1}[j]}\\
 \vdots & \ddots & \vdots \\
{h_{R,\left\lceil {M/2} \right\rceil  + 1}^{j,N}[j]}& \cdots &{h_{R,\left\lceil {M/2} \right\rceil  + N}^{j,N}[j]}
\end{array}} \right]}_{{\bf{\hat H}}_R^j[t]}}^{ - 1}}}
{ \times \left[ {\begin{array}{*{20}{c}}
{\sum\limits_{m = N + 1}^{\left\lceil {M/2} \right\rceil } { - \bar h_{1,m}^{j,1}[j]{s_m}} [j]}\\
 \vdots \\
{\sum\limits_{m = N + 1}^{\left\lceil {M/2} \right\rceil } { - \bar h_{1,m}^{j,N}[j]{s_m}} [j]}
\end{array}} \right],}
\end{array}
\end{small}
\end{equation}
\begin{equation}
\label{eq41}
\begin{small}
\setlength{\arraycolsep}{0.5pt}
\underbrace {\left[ {\begin{array}{*{20}{c}}
{{x_{\left\lceil {M/2} \right\rceil  + N + 1}}[j]}\\
 \vdots \\
{{x_{\left\lceil {M/2} \right\rceil  + 2N}}[j]}
\end{array}} \right]}_{{\bf{\hat x}}_R^{j + 1}[j]}{\rm{ = }}{\underbrace {\left[ {\begin{array}{*{20}{c}}
{h_{R,\left\lceil {M/2} \right\rceil  + N + 1}^{j + 1,1}[j]}& \cdots &{h_{R,\left\lceil {M/2} \right\rceil  + 2N}^{j + 1,1}[j]}\\
 \vdots & \ddots & \vdots \\
{h_{R,\left\lceil {M/2} \right\rceil  + N + 1}^{j + 1,N}[j]}& \cdots &{h_{R,\left\lceil {M/2} \right\rceil  + 2N}^{j + 1,N}[j]}
\end{array}} \right]}_{{\bf{\hat H}}_R^{j + 1}[j]}}^{ - 1}
 \times \left[ {\begin{array}{*{20}{c}}
{\sum\limits_{m = \left\lceil {M/2} \right\rceil {\rm{ + }}N + 1}^M { - \bar h_{2,m}^{j + 1,1}[j]{s_m}} [j]}\\
 \vdots \\
{\sum\limits_{m = \left\lceil {M/2} \right\rceil {\rm{ + }}N + 1}^M { - \bar h_{2,m}^{j + 1,N}[j]{s_m}} [j]}
\end{array}} \right],
\end{small}
\end{equation}
where ${{\bf{\hat H}}_R^j[t]}$ and ${{\bf{\hat H}}_R^{j+1}[j]}$ are the channel matrices between relay $R$'s directional antennas and users' antennas. The received signals of user $j$ and $j+1$ are changed to (\ref{eq42}) and (\ref{eq43}) respectively,
\begin{equation}
\label{eq42}
\begin{small}
\setlength{\arraycolsep}{0.5pt}
 {{{\boldsymbol{\hat y}}}_j}[j]= {{{\boldsymbol{\bar y}}}_j}[j] + {\bf{\hat H}}_{R}^{j}[j]{\boldsymbol{\hat x}}_{R}^{j}{{[j]}}
= \left[ {\begin{array}{*{20}{c}}
{\bar h_{1,1}^{j,1}[j]}& \cdots &{\bar h_{1,N}^{j,1}[j]}\\
 \vdots & \ddots & \vdots \\
{\bar h_{1,1}^{j,N}[j]}& \cdots &{\bar h_{1,N}^{j,N}[j]}
\end{array}} \right]\left[ {\begin{array}{*{20}{c}}
{{s_1}[j]}\\
 \vdots \\
{{s_N}[j]}
\end{array}} \right],
\end{small}
\end{equation}
\begin{equation}
\label{eq43}
\begin{small}
\setlength{\arraycolsep}{0.5pt}
{{{\boldsymbol{\hat y}}}_{j+1}}[j] = {{{\boldsymbol{\bar y}}}_{j+1}}[j] + {\bf{\hat H}}_R^{j+1}[j]{\boldsymbol{\hat x}}_R^{j+1}[j]
=\left[ {\begin{array}{*{20}{c}}
{\bar h_{2,\left\lceil {M/2} \right\rceil  + 1}^{{j+1},1}[j]}& \cdots &{\bar h_{2,\left\lceil {M/2} \right\rceil  + N}^{{j+1},1}[j]}\\
 \vdots & \ddots & \vdots \\
{\bar h_{2,\left\lceil {M/2} \right\rceil  + 1}^{{j+1},N}[j]}& \cdots &{\bar h_{2,\left\lceil {M/2} \right\rceil {{ + N}}}^{{j+1},N}[j]}
\end{array}} \right]\left[ {\begin{array}{*{20}{c}}
{{s_{\left\lceil {M/2} \right\rceil  + 1}}[j]}\\
 \vdots \\
{{s_{\left\lceil {M/2} \right\rceil  + N}}[j]},
\end{array}} \right]
\end{small}
\end{equation}
where ${{{\boldsymbol{\hat y}}}_j}[j]$ and ${{{\boldsymbol{\hat y}}}_{j+1}}[j]$ can be decoded within slot $j$.

In the second phase, DRIR algorithm is adopted in slot $t=\varphi + 1$ to compensate eliminated symbols of all users in previous slots. Specifically, relay $R$ utilizes moderate delayed CSI to design regenerated interference signals as
\begin{equation}
\label{eq44}
\begin{split}
{\boldsymbol{\hat x}}_R^{j + 1}[\varphi {\rm{ + 1}}] = {\bf{\hat H}}_R^{j + 1}{[\varphi {\rm{ + 1}}]^{ - 1}} \times & [{s_{N + 1}}[j],  \ldots ,{s_{\left\lceil {M/2} \right\rceil }}[j],{s_{\left\lceil {M/2} \right\rceil  + N + 1}}[j + 1], \\
& \ldots ,{s_M}[j + 1],{s_{N + 1}}[j + K], \ldots ,{s_M}[j + (\varphi  - 1)K]]_,^T
\end{split}
\end{equation}
where ${\boldsymbol{\hat x}}_R^{j + 1}[\varphi {\rm{ + 1}}]$  is the regenerated interference signals from relay $R$ to the user $j+1$ with directional antennas. Due to the characteristic of directional antennas, there is no additional IUIs generated in network. Therefore, the received signals of user $j+1,\forall j+1 \in {\boldsymbol K}$ are
\begin{equation}
\label{eq45}
{{\boldsymbol{\hat y}}_{j+1}}{\rm{[}}\varphi {\rm{ + 1] = }}{\bf{\hat H}}_R^{j+1}[\varphi {\rm{ + 1}}]{\boldsymbol{\hat x}}_R^{j+1}{\rm{[}}\varphi {\rm{ + 1]}}.
\end{equation}
From (\ref{eq45}), the missing symbols of user $j+1$ in previous slots get compensated, i.e., ${s_{N + 1}}[j], \ldots ,$ ${s_{\left\lceil {M/2} \right\rceil }}[j],{s_{\left\lceil {M/2} \right\rceil  + N + 1}}[j + 1], \ldots ,{s_M}[j + 1],{s_{N + 1}}[j + K], \ldots ,{s_M}[j + (\varphi  - 1)K]$. Herein, the proposed scheme decodes part of desired symbols in real time, and meanwhile, obtains the rest of missing symbols at last.

\subsection{DoD and DoF of HAA-CIE-RIR Scheme}
In this subsection, we analyze the DoD and DoF of proposed two  schemes for the considered $K$-user $M \times N$ BC networks. Assume that the configuration is set to $(K,M,N)$ and the data set consists of $A$ symbols with priority $1$ and $B$ symbols with priority $5$.

For HAA-CIE-RIR, according to the relationship between $M$ and $N$, the total number of time slots is divided into two cases. If $\left\lceil {M/2} \right\rceil {{ \ >  N}}$, two phases exist and each of them occupies $\varphi=K\left\lfloor {{N \mathord{\left/{\vphantom {N {(M - 2N)}}} \right. \kern-\nulldelimiterspace} {(M - 2N)}}} \right\rfloor $ slots and $1$ slot, respectively. In each slot of the first phase, $2N$ symbols get decoded and, in last slot of the second phase, $KN$ symbols get decoded. Therefore, the scheme experiences $\varphi  + 1$ slots and obtains a total of $2N\varphi+KN$ expected symbols, which is partially superior to that of HAA-PIE-RIR scheme. If $\left\lceil {M/2} \right\rceil {{   \le   N}}$, there is only the first phase taking $\varphi$ slots. From which, $M$ symbnols get decoded in each slot. Thus, the proposed scheme experiences $\varphi$ slots and obtains $\varphi M$ symbols in all, which keeps same performance as HAA-PIE-RIR scheme locally. Note that, if $\left\lceil {M/2} \right\rceil > N$, we assume that  $A+B$ is divisible by $2N\varphi+KN$ and, if $\left\lceil {M/2} \right\rceil { \le  N}$, we assume that $A+B$ is divisible by $\varphi M$. Herein, the DoF and DoD of HAA-CIE-RIR scheme are derived as (\ref{eq46}) and (\ref{eq47}), respectively:
\begin{equation}
\label{eq46}
\begin{small}
DoF_{HAA-CIE-RIR} = \left\{ {\begin{array}{*{20}{c}}
M&{\left\lceil {M/2} \right\rceil \; \le N}\\
{(2N\varphi  + KN)/(\varphi  + 1)}&{\left\lceil {M/2} \right\rceil \; > N} \end{array}} \right.
,
\end{small}
\end{equation}
\begin{equation}
\begin{small}
\label{eq47}
DoD_{HAA-CIE-RIR} = \left\{ {\begin{array}{*{20}{c}}
\begin{array}{l}
\frac{{4(\varphi  + 1){A^2}}}{{10(2N\varphi  + KN)(A + B)}} + \frac{{(\varphi  + 1)(A + B)}}{{10(2N\varphi  + KN)}} - \frac{{(5A + B)(\varphi  + 1)}}{{10(A + B)}}
\end{array}& {\left\lceil {M/2} \right\rceil \; > {{N}}}\\
{\frac{{4{A^2}}}{{10M(A + B)}} + \frac{{A + B}}{{10M}} - \frac{{5A + B}}{{10(A + B)}}}& {\left\lceil {M/2} \right\rceil \; \le {{N}}}
\end{array}} \right.
.
\end{small}
\end{equation}

Compared with the RIA and the HAA-IPIE-RIR schemes, the HAA-CIE-RIR scheme can be regarded as a trade-off tactic, i.e., it sacrifices part of DoF gain to reduce computational complexity. On the one hand, the HAA-CIE-RIR scheme is roundly superior to the HAA-PIE-RIR scheme in terms of DoD and DoF, while it brings computaional burden at the relay side. On the other hand, the HAA-CIE-RIR scheme is more attractive than the HAA-IPIE-RIR scheme regarding lower computational cost but, due to the waste of time resources, both DoD and DoF of the former is locally inferior to the later. The details are analyzed in Section VI.

\section{Numerical Results}
In this section, numerical results are provided to verify the achievable performance of the proposed schemes. Specifically, we take three kinds of performance metrics, i.e., DoF, DoD and the computational cost, and meanwhile, we introduce two benchmark schemes, i.e., TDMA scheme and RIA scheme. Then, for $K$-user MIMO BC networks, we set the configuration as $(K,M,N)$. Note that applicable scopes of our proposed schemes are quite different. Thus, we take transceiver antennas ratio $\rho$ to distinguish them and its definition is
\begin{equation}
\label{eq48}
\rho  = {M \mathord{\left/{\vphantom {M N}} \right.\kern-\nulldelimiterspace} N}.
\end{equation}
From (\ref{eq48}), on the one hand, the $\rho$ of RIA scheme ranges from 1 to 2 and its DoF has been saturated when $\rho=2$. This means that the value of DoF no longer increases with the continued growth of $\rho$. However, for our proposed schemes, when the value of $\rho$ changes from $1$ to $2$, the performance measured by DoF and DoD is superior than that of the RIA scheme, and meanwhile, when $\rho$ is greater than 2, DoF and DoF can continue to be optimized by our proposed schemes. On the other hand, for our proposed schemes, there are distinct critical points caused by round operation. For instance, as for the HAA-PIE-RIR scheme, the best improvement effect can be obtained when $N$ is divisible by $\left\lceil {M{{/}}K} \right\rceil - N$. However, as for the HAA-CIE-RIR scheme, the condition is replaced with another condition that $N$ is divisible by $M-2N$. Thus, the choice of parameters, i.e., $M$, $N$ and $A+B$, is difficult to meet all restrictions. In order to meet as many conditions as possible, we set configuration of networks as $(3,60,N)$, change the value of $N$ from $10$ to $25$, and take $A=1400,B=1400$. In this case, the proposed schemes compare with each other on three aspects, i.e., DoF, DoD and computational cost, where corresponding simulation results are presented as Fig. 6, Fig. 7 and Fig. 8, respectively. Finally, we provide suggestions about pratical application scenarios for each proposed schemes.

\subsection{Degree of Delay}
We first compare the DoF performance of our proposed schemes with that of benchmark schemes.  From Fig. 6, we observe that all curves rise with the increasing number of receiving antennas per user $N$. This is attributed to an increase of space resources. Nonetheless, we still find that, when $N=20$, the DoF of HAA-IPIE-RIR scheme keeps constant, i.e., $60$, and the DoF of HAA-CIE-RIR scheme slumps into $45$. For the former, it is due to the fact that the value of $N$ reaches the critical point at $KN=M$. Since then,  the DoF of HAA-IPIE-RIR scheme mainly depends on transmitting antennas $M$ while $M$ is fixed, in other words,  the DoF of HAA-IPIE-RIR scheme is saturated in our considered case. For the latter, it reflects the loss of DoF caused by round operations vividly. In addition, compared with the RIA scheme, our proposed schemes achieve higher DoF gains throughout the whole process of $M$'s variation. Specifically, on the basis of RIA scheme, HAA-PIE-RIR scheme, HAA-CIE-RIR scheme, and HAA-IPIE-RIR scheme ulteriorly improve the performance of DoF to $50\%$, $150\%$ and $150\%$ as $N=10$, respectively, and at the same time, obtain a higher DoF gain of $50\%$, $64.3\%$ and $90\%$ when $N=25$, respectively. This phenonmenon lies in fact that  HAA-PIE-RIR scheme utilizes space resources more adequately than RIA scheme in the first phase and meanwhile, HAA-CIE-RIR scheme and HAA-IPIE-RIR scheme obtain spatial gain brought by multi-users.
\begin{figure}[h]
\vspace{-1.0em}
\centering
\includegraphics[width=2.5in]{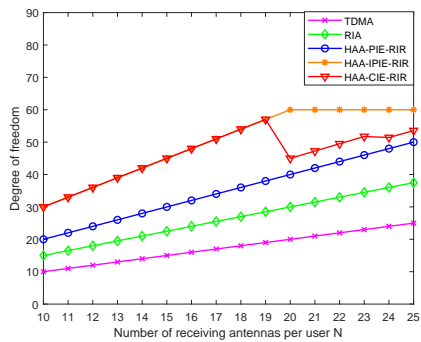}
\caption{The DoF of three proposed schemes.}
\label{Fig6}
\end{figure}
\vspace{-1.0em}

\subsection{Degree of Freedom}
With our proposed schemes, Fig. 7 reflects the improvement effects on the problem of information delay. In general, when the number of receiving antennas increases, DoDs of all schemes get decreased. This is because the number of symbols that can be decoded within one slot gets raised, which means that the total number of slots is cut down and corresponding DoD becomes lower. With further analysis, three proposed schemes have different degrees of reducing DoD. Specifically, compared with TDMA scheme, RIA scheme reduces DoD by the improvement of DoF which leads to a lower amount of slots. Based on it, three proposed schemes further relieve the problem about information delay. Specifically, HAA-PIE-RIR scheme utilizes hybrid antenna array structure of relay to achieve timely decoding of desired symbols in the first phase and eliminate IUI of all users simultaneously. Under the effect of relay, the problem of information delay in one period can be alleviated, and meanwhile, the utilization of space resources is further improved. By quantifying the extent of improvement with DoD, it indicates that, compared with RIA scheme, DoD is further reduced to $31.0\%$ when $N=10$ and DoD is cut down to $38.9\%$ as $N=25$. As for HAA-IPIE-RIR scheme and HAA-CIE-RIR scheme, the more space resources from multi-users are exploited, so that DoD becomes lower than that of the RIA scheme, where the DoD gets decreased to $53.6$ and $56.5$ when $N=10$ and $N=19$. Note that, when $N$ ranges from $20$ to $25$, the performance of HAA-CIE-RIR scheme deteriorates. This lies in the fact that HAA-CIE-RIR scheme has reached one critical point while the others have not. Thus, only the HAA-CIE-RIR scheme experiences DoF loss and its corresponding total number of slots increases, leading to a higher DoD. Nevertheless, the performance of HAA-CIE-RIR scheme is still better than that of RIA scheme.
\begin{figure}[h]
\vspace{-1.0em}
\centering
\includegraphics[width=2.5in]{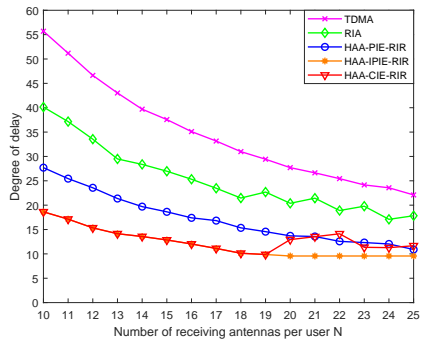}
\caption{The DoD of three proposed schemes.}
\label{Fig7}
\end{figure}
\vspace{-1.0em}

\subsection{Computational Complexity}
For our proposed schemes, they have one additional advantage, i.e., the burden of computational complexity migrates from users to the relay. Referring to \cite{aslan2019trade}, for one node, we take the maximum times of multiplication within unit time as the computational cost. Thus, the computational cost mainly depends on the step whose computational complexity is the highest. Thus, the computational cost of IA schemes is formulated as shown in Table II.
\begin{table}[h]
\centering
\renewcommand\arraystretch{0.5}
\setlength\tabcolsep{1pt}
 \centering
 \setlength{\abovecaptionskip}{0.cm}
 \setlength{\belowcaptionskip}{-0.cm}
 \caption{ computational complexity of IA schemes }
\begin{small} 
\begin{tabular}[b]{cccccccc}
\cr \toprule[1.2pt]
Scheme & Object & Complexity & Condition& Scheme  & Object & Complexity & Condition
\cr \toprule[1.2pt]
\cr RIA & Relay & - & ${M \le 2N}$ & HAA-PIE-RIR & Relay & ${2\left\lfloor {M/2} \right\rfloor {N^5}}$ & ${\left\lceil {M{\rm{/2}}} \right\rceil \le N}$\\
\cr       &         & - &  ${M > 2N}$ & & & ${(2\left\lfloor {M/2} \right\rfloor {\rm{ + 2)}}{N^5}}$ & ${\left\lceil {M{\rm{/2}}} \right\rceil > N}$\\
\cr       & User & ${{{\left\lfloor {{N \mathord{\left/
{\vphantom {N {(M - N)}}} \right.
 \kern-\nulldelimiterspace} {(M - N)}}} \right\rfloor }^3}{M^3}}$ & ${M \le 2N}$ & & User & ${{{\left\lceil {M{\rm{/2}}} \right\rceil }^3}}$ & ${\left\lceil {M{\rm{/2}}}\right\rceil\le N}$\\
\cr & & ${{M^3}}$ & ${M > 2N}$ & & & ${{N^3}}$ & ${\left\lceil {M{\rm{/2}}} \right\rceil > N}$\\
\cr HAA-IPIE-RIR & Relay & ${{{(KN)}^6}{M^2}}$ & ${\left\lceil {M{{/K}}} \right\rceil \le N}$ &  HAA-CIE-RIR & Relay & ${2\left\lfloor {M/2} \right\rfloor {N^5}}$ & ${\left\lceil {M{\rm{/2}}} \right\rceil \le N}$\\
\cr & & ${{{(KN)}^6}{M^2} + K{N^5}}$ & ${\left\lceil {M{{/K}}} \right\rceil > N}$ & & & ${(2\left\lfloor {M/2} \right\rfloor {\rm{ + }}K{\rm{)}}{N^5}}$ & ${\left\lceil {M{\rm{/2}}} \right\rceil > N}$\\
\cr & User & ${{{\left\lceil {M{{/}}K} \right\rceil }^3}}$ & ${\left\lceil {M{\rm{/}}K} \right\rceil \le N}$ & & User & ${{{\left\lceil {M{\rm{/2}}} \right\rceil }^3}}$ & ${\left\lceil {M{\rm{/2}}} \right\rceil \le N}$\\
\cr & & ${{N^3}}$ & ${\left\lceil {M{\rm{/}}K} \right\rceil > N}$ & & & ${{N^3}}$ & ${\left\lceil {M{\rm{/2}}} \right\rceil > N}$\\
 \cr \bottomrule[1.2pt]
 \end{tabular}
\end{small}
 \label{tab:addlabel}
\vspace{-2.0em}
\end{table}

Subsequently, we simulate the computational complexity of the IA schemes and the results are plotted in log scales. From Fig. 8, it can be seen that part of burden related to computational complexity is migrated from users to the relay. Specifically, for HAA-IPIE-RIR scheme, the computational complexity at the relay side is considerably high, which makes it not practical in massive MIMO scenarios. This is because the complexity of precoding matrix design is too high. As for HAA-IPIE-RIR scheme and HAA-CIE-RIR scheme, although the computational complexities of relay are much higher than that of RIA scheme at user side, when transceiver antennas ratio is comparative large, e.g., $M/N=6$, the increment of computational complexity is tenfold, which is relatively acceptable. In addition, we find that computational complexity of RIA's user keeps constant. This is due to the saturation of DoF where the determinant of computational complexity is from the number of transmitting antennas $M$, while $M$ is a constant. Thus, we compare three proposed schemes with RIA scheme at user side. It is noticed that, when $N=10$, computational complexity of RIA's user is $1000$ times than that of proposed schemes' user, and at the same time, when $N=25$, computational complexity of RIA's user is $100$ times than that of three proposed schemes' user. It means that users' burden of computational cost is reduced with our proposed schemes.
\begin{figure}[h]
\vspace{-1.0em}
\centering
\includegraphics[width=2.5in]{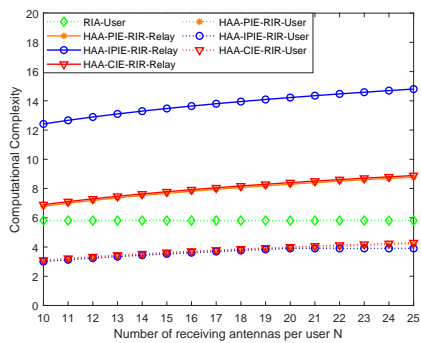}
\caption{The computational cost of three proposed schemes in log scales.}
\label{Fig8}
\end{figure}
\vspace{-1.0em}

\subsection{Application Scenarios}
In practice, HAA-IPIE-RIR scheme is more suitable for the case that the number of antennas at the receivers is relatively small, e.g., the receivers are mobile users, or the relay has strong computing capability. In these scenarios, HAA-IPIE-RIR scheme can obtain the optimal performance in DoD and DoF. However, for HAA-CIE-RIR scheme, it is more applicable for the case that the number of antennas at the receivers is large, e.g., the receivers are the base stations. With HAA-CIE-RIR scheme, the computing burden of the relay will be significantly relieved. Besides, HAA-PIE-RIR scheme does not need to consider the hardware cost at the relay for frequent beam switching or the computing burden for precoding, while it is mainly used for the case with $2$-user. If we adopt HAA-PIE-RIR scheme in $K$-user application scenarios, its performance will be worsen than that of HAA-IPIE-RIR scheme and HAA-CIE-RIR scheme.

\section{Conclusion and Future Work}
For interference networks, we first propose DoD to quantify the issue of information delay. It comprehensively models and captures the issue about time delay of transmission schemes and time-sensitive services. Then, we simulate the DoD of three typical schemes, i.e., TDMA scheme, BD-TDMA scheme, and RIA scheme, to show that its main influence factors are delay sensitive factor, size of data set, and  queueing delay slot. Among them, the first two can be regarded as weighting factors of DoD which are not related to  the transmission schemes. To improve DoD, we consider reducing the queueing delay slot, and design three novel joint IA schemes for BC networks with different number of users, i.e., HAA-PIE-RIR scheme, HAA-IPIE-RIR scheme, and HAA-CIE-RIR scheme. Based on the first scheme, the second scheme extends the application scenarios from $2$-user to $K$-user while brings huge computational burden. Meanwhile, the third scheme relieves such a computational complexity burden but it leads to certain DoF loss. Simulation results show that the HAA-CIE-RIR scheme has advantage in terms of lower computational complexity and the HAA-IPIE-RIR scheme achieves the higher DoF and the lower DoD. Meanwhile, the performances of the three proposed schemes are roundly superior than that of the RIA scheme. For the future work, we will apply DoD to other IA schemes to study the issue of information delay. Meanwhile, we will consider finite accuracy CSI in some specific scenarios which cannot ensure the correctness of channel estimation.

\ifCLASSOPTIONcaptionsoff
  \newpage
\fi

\end{document}